\documentclass[letterpaper]{article} % DO NOT CHANGE THIS
\usepackage{aaai2026}  % DO NOT CHANGE THIS
\usepackage{times}  % DO NOT CHANGE THIS
\usepackage{helvet}  % DO NOT CHANGE THIS
\usepackage{courier}  % DO NOT CHANGE THIS
\usepackage[hyphens]{url}  % DO NOT CHANGE THIS
\usepackage{graphicx} % DO NOT CHANGE THIS
\usepackage{natbib}  % DO NOT CHANGE THIS AND DO NOT ADD ANY OPTIONS TO IT
\usepackage{caption} % DO NOT CHANGE THIS AND DO NOT ADD ANY OPTIONS TO IT
\usepackage{algorithm}
\usepackage{algorithmic}
\usepackage{booktabs}
\usepackage{subcaption}
\usepackage{newfloat}
\usepackage{listings}
\DeclareCaptionStyle{ruled}{labelfont=normalfont,labelsep=colon,strut=off} % DO NOT CHANGE THIS
\floatstyle{ruled}
\newfloat{listing}{tb}{lst}{}
\floatname{listing}{Listing}
\usepackage{xcolor} 
\usepackage{enumitem}
\usepackage{quoting}
\quotingsetup{vskip=1pt}

\newcommand{\red}[1]{\textcolor{black}{#1}}

\title{Studying People to Study AI: Expert Perspectives on the Epistemic Fit and Barriers of Human Research in AI Safety \& Ethics}
\author{
    Jessica Y. Bo\textsuperscript{\rm 1}, Paula Akemi Aoyagui\textsuperscript{\rm 1}, Shalaleh Rismani\textsuperscript{\rm 2}, Dipto Das\textsuperscript{\rm 1}, Syed Ishtiaque Ahmed\textsuperscript{\rm 1}, \\ Ashton Anderson\textsuperscript{\rm 1}
}
\affiliations{
    \textsuperscript{\rm 1}University of Toronto\\\textsuperscript{\rm 2}McGill University \& Mila Quebec AI Institute
}
\usepackage{bibentry}
\begin{document}

\maketitle

\begin{abstract}

Safety risks of AI are becoming increasingly evident in human interactions with AI technologies. 
The prominent approaches to evaluating these risks favor technical methods, such as benchmarks and AI simulations, \red{often sidelining} empirical research \red{with human subjects.} 
% despite the latter's established role in grounding claims about real-world harm. 
To examine this \red{apparent gap in the acceptance of human research}, we conduct an expert survey ($n = 93$) and expert interviews ($n = 17$) with AI Safety \& Ethics (AISE) researchers from Technical, Sociotechnical, Governance, and Normative backgrounds. 
Our findings suggest that although there is a consensus that human research is valuable for generating evidence for AISE, its adoption and acceptance are constrained by perceived validity issues, tangible resource barriers, epistemic and personal preferences in methods, and infrastructural constraints from the broader research community.  
In particular, Technical researchers tend to value human research less and collaborate across disciplines less, \red{suggesting an epistemic tension towards human methods}.
We propose recommendations for establishing the epistemic fit of human research within AISE and bridging the prohibitive limitations that researchers face, \red{while avoiding performative `human-washing'}. 

\end{abstract}


\section{Introduction}

Research on the safe and ethical design and deployment of AI systems is a multi-faceted, fast-evolving, and critical area of AI research, encompassing a spectrum of disciplines, both technical and sociotechnical in nature \cite{leslie2019understanding, hendrycks2025introduction}. As the boundaries of the research community are continuously evolving, two dominant yet divisive camps within the broader field emerge as AI Safety (AIS) and AI Ethics (AIE) \cite{Gyevnar2026-uq}. In this paper, we group them together with other adjacent disciplines, referring to the broader community as \textit{AISE} for their shared concern about the impacts of AI on humans and society. And yet an interesting paradox remains unaddressed: despite AISE's mission to anticipate and prevent harm to humans, empirical human-centred research methodologies\footnote{\red{We adopt the broad definition of human empirical research as direct data collection from people (e.g., interviews, surveys, behavioral experiments, fieldwork, participatory design, social media analysis) to study how AI systems affect individuals and society. }} are \red{still lacking} in the field. 

Specifically, we highlight that AISE research involving any human subjects is rare within published literature \cite{weidinger2023sociotechnical}, being outnumbered by technical or normative approaches like LLMs simulations of human reactions to AI outputs \cite{li2026llm, ni2026survey}, static benchmarks with binary labels for harm \cite{reuel2024betterbench, yu2026should}, and AI constitutions guided by assumptions about normative value alignments \cite{bai2022constitutional}. These approaches sacrifice construct validity for cost and time efficiency, risking the failure to capture the desired safety construct \cite{bean2026measuring}. Therefore, the evidence of harm for AISE is not well-substantiated if researchers rely on LLM simulations, normative assumptions, or incidental case studies. 
% \red{Beyond this evidence gap, there also a commensurability gap in how human research is accepted and taken up as legitimate, where evidence generated under different epistemologies is not always treated as equal.}
% Research performed with human participants can fill the evidence gap \cite{weidinger2023sociotechnical} \red{reframe around acceptance and commensurability, whose evidence counts?}. 
While there is a recent uptick in human subject research around interactional and experiential harms of AI --- like large-scale experiments on emotional reliance \cite{kirk2025neural, fang2025ai} and sycophancy \cite{cheng2026sycophantic, rathje2025sycophantic} --- the extent to which human methods fit into the evolving \red{epistemic boundaries} of AISE is still unclear. The International AI Safety Report \cite{bengio2025international}, a review of AI risks for governance decision-making, calls for better evidence via observational records of the scale of harm from deepfakes and experimental studies on the impact of AI manipulation.

While past works \cite{Gyevnar2026-uq, roytburg2025mind} illuminated differences between research communities within AISE via literature reviews, we focus on the researchers themselves to understand: (\textbf{RQ1}) \textit{across AISE disciplines, what do experts see as the epistemic fit of human research?}; and (\textbf{RQ2}) \textit{What are the barriers that AISE researchers face in performing human research?} Importantly, while literature surveys analyse published research, we directly capture experts' motivations and deterrents behind \textit{how} the research is pursued, or \textit{why} it's not \red{performed or valued} \cite{agapie2022using,paskov2026rcts}. \textbf{To understand these lived experiences, we surveyed ($n = 93$) and interviewed ($n = 17$) experts from \textit{Technical}, \textit{Sociotechnical}, \textit{Governance}, and \textit{Normative} disciplines who are actively contributing to the field.}

\red{We find both a gap in evidence (lack of human research performed due to structural barriers) and a gap in commensurability between human-centered and technical methods}. 
While experts across disciplines agree that more evidence of AI's impact on humans is needed, and that technical-only research is often plagued with construct validity issues, they hold varied opinions on which human methods generate valid evidence. These differences are largely driven by epistemic incompatibility and low familiarity with human subject methodologies. \textit{Technical} researchers are more likely to undervalue the contributions of human methods, and to collaborate across disciplines less frequently. Qualitative and mixed-methods approaches stand out as an underutilized yet highly valued means of explaining mechanisms of harm, which challenges the quantitative leaning of AISE. In terms of barriers, participants are foremost impacted by a lack of tangible research resources like time, funding, and participant access. Many also report tensions arising from their mentors, institutions, and sectors. In the following sections, we present relevant literature, describe our data collection and analysis methods, detail our results, and discuss the implications for cross-disciplinary and cross-sector collaboration --- including that AISE community should critically, not performatively, \red{recognize and legitimize} human research to generate evidence for harms from AI.

\section{Related Works}
We position our contributions in relation to a body of computing research that concerns epistemic divides in multi-disciplinary areas, the evolution of AISE and its sub-communities, and the problems where human research has been applied (or has the opportunity to be) in AISE.

\subsection{Epistemic Disciplinary Divides}
AISE is far from the first research community to grapple with the tensions inherent in uniting practitioners from disparate disciplinary traditions. Parallel challenges have unfolded in computational healthcare \cite{agapie2022using}, environmental science \cite{miller2008epistemological}, and other complex, problem-driven fields that present tensions in the choice of methods to use, frameworks to ground in, and problems to work on. \citeauthor{kuhn1970structure} describes the nature of scientific progress as waves of research that fundamentally reframes the accepted paradigm of the scientific community. As fields with multi-disciplinary compositions mature and shift in their epistemologies, it raises questions about what forms of evidence count as legitimate and whose methods should be used to generate knowledge \cite{jacobs2009interdisciplinarity, talbi2025reflections}. Such epistemic boundaries reflect the values embodied by the field, which may be dominated by the majority voices, such as machine learning favouring technical and quantitative optimization over societal impacts \cite{birhane2022values}. 

Human-computer interaction (HCI) also offers a parallel precedent, where the field originated from the positivist and quantitative paradigms of computer science and engineering \cite{duarte2016revisiting, harrison2007three}. Later waves integrated cross-disciplinary theories from cognitive science and phenomenology, but due to the origins of the field being from technical disciplines, interpretivist and qualitative methods are often sidelined or misinterpreted \cite{soden2024evaluating}. Deliberate advocacy for the inclusion of such methods was needed to push the field towards acceptance, offering a roadmap for how AISE may navigate incorporating human research. This aligns with \citeauthor{miller2008epistemological}'s concept of \textit{epistemological pluralism}, which argues that different `ways of knowing' can be integrated to gain a fuller description of a field. In this study, we examine how AISE experts currently navigate the 
clashing paradigmatic tensions of different disciplines, and how human research is positioned to contribute value to the field.

\subsection{Evolution of AI Safety \& Ethics}
%\red{\cite{ahmed2023building} this reference for the epistemic community of AIS might be useful }

% both fields were substantially formalized from mid 2010s onwards as AI systems were being adopted in practice. Their evolution mirrors broader pattersn safety engineering where the focus shifts from technical analysis of components to examining users and systems impacted by these technologies \cite{Swuste2021-qr}. 

%the industrial revolution, where early safety engineering focused on technical failures and component-level analysis, and it was only after major incidents that researchers began drawing on social sciences to study what had gone wrong, and this led to the rise of fields of human factor engineering and system safety \cite{Swuste2021-qr}. AI safety and AI ethics reflect a similar divison. 
While the history of AI Safety and AI Ethics can be traced back to technology ethics \cite{Vallor2016}, science and technology studies \cite{Bijker1987}, and safety engineering \cite{Perrow1984, Rasmussen1997, Swuste2021-qr}, the two fields were substantially formalized from the mid-2010s as AI systems were being adopted in practice. AIS's initial focus was on AI alignment \cite{Russell2019, Amodei2016-eg}---ensuring AI systems behave in accordance with human intentions, often expressed mathematically---with intellectual roots in Bostrom and Yudowsky's existential risk framing of Artificial General Intelligence \cite{Bostrom2014, Yudkowsky2008}. Scholars criticized AIS's ties to effective altruism, longtermism, and rationalism for fostering ideological homogeneity at the expense of excluding non `in-group' voices \cite{Ahmed2023-cr, gebru2024tescreal}. Furthermore, AIS' technology-centric perspective led to the sidelining of complex societal impacts of AI \cite{Dahlgren-Lindstrom2025-ah, Walker2024-cg}. In response to these critiques, AIS has expanded to include broader perspectives such as technical AI governance \cite{Reuel2024-eo} and system safety \cite{Rismani2023-xt, Dobbe2022-ql}. 

In contrast, AIE emerged from longer traditions of studying the societal impacts of algorithmic systems and gained significant momentum when the real-world deployment of AI systems began producing visible, documented harms \cite{Birhane2022-vm, Shelby2023-to}. This prompted research and policy communities to shape high-level ethical principles \cite{Jobin2019-kt} and venues like AIES (AAAI Conference on AI, Ethics, and Society) and FAccT (ACM Conference on Fairness, Accountability, and Transparency) gained prominence. AIE coalesced around a cluster of ethical principles, including fairness, transparency, privacy, sustainability, and accountability, distinctively centering the experiences of impacted communities \cite{CostanzaChock2020} with various levels of rigour and success. %maybe add a phrase like with var

Some scholars working across both communities have increasingly tried to map and close the gap between them \cite{kasirzadeh2025two, shen2024towards, Lazar2023-du, roytburg2025mind}. \citeauthor{Gyevnar2026-uq} identify four common forms of engagement in AI ethics and safety researchers, including radical confrontation, disengagement, compartmentalized coexistence, and critical bridging. They argue for creating spaces that allow for productive exchanges between the two fields. 

\subsection{Human Research in AI Safety \& Ethics}
As AI systems become more deeply integrated into consequential domains, taxonomies of AI risk have increasingly attended to harms that are experienced by people. Sociotechnical frameworks of AI risk \cite{weidinger2022taxonomy, Shelby2023-to} tackle the interactional and experiential dimensions of harm where the user experiences the outputs of the AI via direct interactions, or by existing in an AI-driven society. These include cognitive offloading and skill erosion \cite{chalkidis2026brainrot}, emotional over-reliance on AI systems \cite{saracini2025techno}, gradual disempowerment through AI-mediated decision environments \cite{kulveit2025gradual}, susceptibility to AI-enabled manipulation \cite{matz2024potential}, and much more. 
These harms are emergent, but not hypothetical. Yet dominant risk frameworks continue to foreground technical problems in alignment, robustness, and security \cite{hendrycks2021unsolved}, leaving many of the most immediate human-centered harms underserved \cite{weidinger2023sociotechnical}.

Where human methods do appear in AISE, they tend to be large-scale, quantitative, and narrowly evaluative -- representing the perspectives and behaviours of an \textit{average person} rather than understanding lived experience of harm. Reinforcement learning from human feedback (RLHF) represents the field's most institutionalized form of human engagement for value elicitation \cite{ouyang2022training}, but the human is treated as a training signal instead of the subject of research. Red-teaming \cite{perez2022red} and scalable oversight research \cite{bowman2022measuring} involve human participants but treat them as parts of the evaluation pipeline. Randomized controlled trials (RCTs) testing the impact of AI design on perceptions, beliefs, and performance are becoming more common \cite{paskov2026rcts}, but remain rare due to the high costs involved. In contrast, qualitative, participatory, and mixed-methods approaches have a more established presence in AIE research, in investigating empirical accounts of how AI impacts affected communities \cite{Birhane2022-vm}. However, this methodological tradition has developed largely in parallel to AIS \cite{Gyevnar2026-uq, roytburg2025mind}, keeping the two disciplines effectively separated. 

This methodological narrowness draws parallels to the field of explainable AI (XAI), a case study in which a field built technical methods premised on the assumption that model transparency improves human-AI collaboration, only for empirical studies to reveal that explanations can increase over-reliance and worsen decision outcomes \cite{kaur2024interpretability, bansal2021does}. AISE risks repeating this sequence at a larger scale, investing in technical mitigations for harms whose human dimensions have not yet been adequately characterized. We therefore approach this paper as a way to understand \textit{why the deeply human problems of AISE are not always approached with human research}.

\section{Methods}
We gather AI Safety \& Ethics expert insights through two complementary methods: a survey ($n=93$), and in-depth interviews ($n=17$). This mixed-methods design allows us to identify trends across disciplines and enrich them with nuanced accounts from individual researchers \cite{creswell2017designing, tashakkori2021sage}.

\subsection{Expert Survey}
The survey covers both participants' personal practices and broader perspectives on their discipline and AISE as a field. Inclusion criteria were intentionally broad, requiring only that participants self-identify as working on AISE-relevant research. This encompasses independent researchers without institutional affiliation, governance practitioners who may not publish, and those who may feel excluded by narrower definitions of AI Safety. Professional identities were verified through self-reported areas of focus and, where applicable, professional email addresses. Two attention check questions were included; responses failing both were excluded. The survey comprised four sections (see Appendix
% \footnote{\label{arxiv}\red{See \url{https://arxiv.org/abs/2608.05656} for appendices.}}
\ref{app:survey} for full question text): 
% \red{footnote since we focus on perceptions, the items are not explicitly validated...}

\begin{enumerate}[nosep]
  \item \textbf{Background and Experience:} Professional   background in AISE, including primary research area (\textit{Technical}, \textit{Sociotechnical}, \textit{Governance}, or \textit{Normative}). We use more granular categories than AIS and AIE to be inclusive of disciplines that may fall outside those epistemic boundaries, and further clustering by shared similarities in methods and ontologies.
  \item \textbf{Practice and Aspirations:} Ratings of how often participants currently perform, and how useful they perceive, various human research methods --- \red{we intentionally keep the definition of human research broad to be inclusive of any empirical work with human subjects.}\footnote{\red{However, we noted to participants that research using existing datasets produced by humans do not count.}}
  \item \textbf{Scenario-Based Perceptions:} Ratings of the usefulness of human vs. non-human methods for addressing four AISE risk scenarios, validated to vary on imminence and risk level. This section captures field-level views rather than discipline-specific practices.
  \item \textbf{Barriers to Human Research:} Degree to which resource, knowledge, and methodological barriers have impacted participants' engagement in human research.
\end{enumerate}

\subsection{Expert Interview}
Semi-structured interviews were conducted to gain richer, descriptive accounts to contextualize the survey findings. Each interview covered the participant's research area and its relationship (or lack thereof) to human research, cross-disciplinary collaboration practices, perceived AI harms, and views on the value and fit of human methods in AISE. For those with prior human research experience, we additionally probed barriers they had encountered.

Analysis followed a mixed-methods approach in which survey results provided the organizing structure for qualitative coding \cite{creswell2017designing}. We developed a codebook deductively from the research questions, refined inductively using themes emerging from the interviews \cite{braun2006using}. The initial codebook was developed by the first author and then refined through independent coding by two additional authors, with final themes agreed upon collaboratively by the full team. Author positionality, including expertise in human research and engagement with AISE, is reported at the end of the paper.

\subsection{Recruitment and Participants}
We recruited primarily through the annual conference of the International Association for Safe \& Ethical AI\footnote{\url{https://www.iaseai.org}} (IASEAI 2026), which convenes multidisciplinary AISE researchers across academia, industry, non-profits, and government. Unlike AI Ethics-focused venues such as FAccT and AIES, IASEAI includes a substantial technical AI Safety community, making it well-suited to our cross-disciplinary aims. We contacted attendees from an internal database who met a minimum threshold of research engagement (e.g., at least one published paper or a senior position in a relevant organization). We supplemented this with open recruitment via team social media and targeted Google Scholar searches for authors of high-impact papers in AI safety, ethics, alignment, and harm evaluation — prioritizing underrepresented areas when 
needed.\footnote{For example, to target \textit{Normative}, we contacted recent authors from the \textit{Minds and Machines} journal.} Participation in was not compensated as the experts were motivated to contribute to field-advancing research of their own volition.

The survey sample ($n=93$) comprised primary research areas of \textit{Technical} ($n_T=32$), \textit{Sociotechnical} ($n_S=26$), \textit{Governance} ($n_G=25$), and \textit{Normative} ($n_N=10$). The sample skewed toward WEIRD contexts \cite{septiandri2023weird}, with participants located primarily in North America (47.3\%) and Europe (44.1\%). Most hold or are completing a PhD (67.8\%), work in academia (70.9\%), and are substantially or fully engaged in research (88.2\%). Years of AISE research experience ranged from 1 to 10+ (M=4.44, SD=2.70). Recruiting through IASEAI allowed us to capture non-academic participants while retaining a diverse sample active across sectors. Full demographic details are in Tables \ref{tab:region}--\ref{tab:experience} and Figure \ref{fig:participants} in Appendix
% \footnotemark[2]
\ref{app:survey_participants}.

For interviews, we sought diversity across disciplines and sectors, though participants who opted into a follow-up interview were likely already engaged with human research, introducing selection bias. The final interview sample ($n=17$) included participants across all four research areas (6 \textit{Technical}, 5 \textit{Sociotechnical}, 4 \textit{Governance}, and 2 \textit{Normative}), across experience levels (M=4.7 years), and across academia, non-profits, and industry. See Table \ref{tab:interview} for details.

\begin{table}[t!]
\small
    \centering
    \begin{tabular}{cccc}\toprule
         \textbf{PID}&\textbf{Yrs. of Exp.}&\textbf{Sector \& Role}&\textbf{Region}\\\midrule
         P1$_{S}$& 2& Academia (Student)&N. America\\ 
 P2$_{T}$& 1& Academia (Student)&N. America\\
 P3$_{G}$& 6& Academia (Researcher)&Europe\\
 P4$_{S}$& 5& Academia (Researcher)&N. America\\
 P5$_{T}$& 2& Academia (Student)&Europe\\
 P6$_{S}$& 5& Academia (Researcher)&N. America\\
 P7$_{G}$& 6& Academia (Researcher)&N. America\\
 P8$_{G}$& 9& Non-Profit (Manager)&Europe\\
 P9$_{S}$& 4& Academia (Student)&Europe\\
 P10$_{T}$& 1& Academia (Researcher)&N. America\\
 P11$_{T}$& 9& Industry (Researcher)&N. America\\
 P12$_{T}$& 2& Academia (Student)&N. America\\
 P13$_{N}$& 7& Academia (Researcher)&Europe\\
 P14$_{T}$& 3& Non-Profit (Researcher)&Asia\\
 P15$_{G}$& 10+& Academia (Researcher)&N. America\\
 P16$_{N}$& 6& Academia (Student)&N. America\\
 P17$_{S}$& 2& Academia (Student)&N. America\\ \bottomrule\end{tabular}
    \caption{Summary of interview participants. They are referred to in the text by their participant number \textit{PID$_{Primary\ Area}$}, where the subscript indicates the primary area they reported in the survey (e.g. \textit{S} for \textit{Sociotechnical}). 
    % For the role, `Researcher' can refer to faculty, research leads, postdocs, or research engineers. 
    }
    \label{tab:interview}
\vspace{-0.5em}
\end{table}

\begin{figure*}
    \centering
    \includegraphics[width=\linewidth]{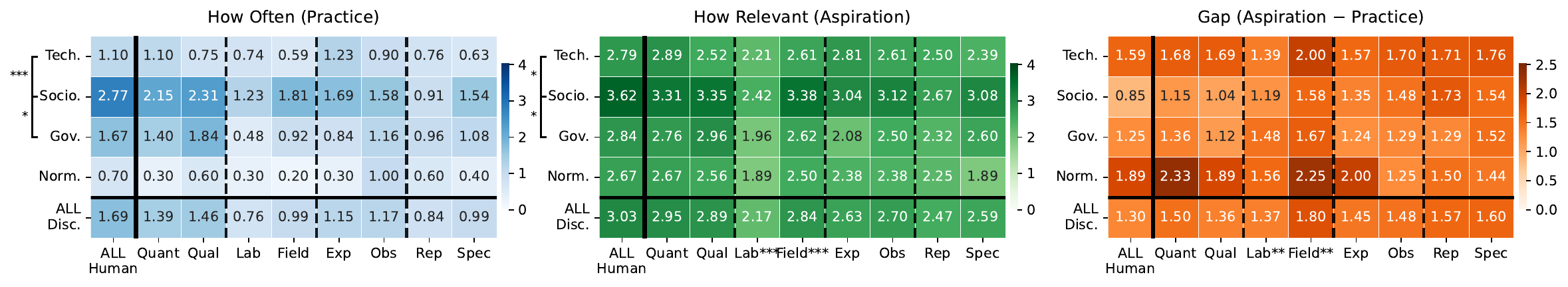}
    \caption{Within their disciplines, how often do participants \textit{practice} human research, and how much they would \textit{aspire} to perform methods relevant to their research. Both are measured on a Likert scale mapping to values between 0 - 4. Disciplines or research dimensions that have a significant difference are indicated (*** for $p<.001$, ** for $p<.01$, * for $p<.05$).}
    \label{fig:practice_aspirations}
\vspace{-0.5em}
\end{figure*}

\section{Results}

We organize our results in two sections: RQ1) how experts perceive the epistemic fit of human research within AISE, and RQ2) the resource, infrastructural, and sector limitations that impede it. Each section is supported by quantitative survey analyses and thematic interview findings. Of the 17 interviewees, the majority had direct or adjacent experience with human research---six had conducted it themselves, two engaged domain experts in participatory ways, and two had prior human research experience in other fields---providing a range of perspectives on both practice and aspiration.

\subsection{RQ1: Epistemic Value and Disciplinary Divides}

We explore the epistemic fit of human research by examining the degree to which human research methods are perceived as generating legitimate, useful evidence within AISE. Specifically, what can human studies offer that other approaches cannot? How do disciplinary training and epistemic priors shape researchers' 
valuations of that evidence? And what does the interplay of disciplines within AISE mean for the acceptance of human methods?

\subsubsection{The Human Evidence Gap.}
Participants agreed that AISE research objectives are becoming increasingly human-centered and value-driven, shifting from theoretical and hypothetical problems toward the interactions between AI systems and people (P6$_{S}$, P9$_{S}$, P12$_{T}$, P15$_{G}$, P17$_{S}$) \cite{dotan2019value, birhane2022values}. High-impact sociotechnical harms demand urgency in collaborative action (P3$_{G}$) and play into the moral conscience of AI researchers (P17$_{S}$). However, dominant computational approaches are frequently limited by \textit{construct validity} problems, as they struggle to capture subjective constructs like what harm to humans actually means in practice (P2$_{T}$, P6$_{S}$, P9$_{S}$, P11$_{T}$, P12$_{T}$, P15$_{G}$). Many technically oriented participants themselves critiqued a positivism-rooted ideology that reduces nuanced phenomena to single optimizable metrics, flattening the very constructs AISE aims to evaluate. Gathering evidence from domain-specific stakeholders can challenge these assumptions and surface unexpected findings (P1$_{S}$, P4$_{S}$), though there is still a lack of consensus on which AISE issues warrant evaluation (P6$_{S}$) or which human methods are appropriate (P14$_{T}$).

 \begin{quoting}
 \footnotesize
     \textit{People have this obsession with a single metric. ``We can just make a scale from 0 to 100" --- it is totally arbitrary. There is no construct validity. It's literally a number we decided, and now we're going to declare this as the metric for goodness, and everyone will align around those numbers. } --- \textbf{P11$_{T}$}, on the construct validity of benchmarks.
 \end{quoting}
 % \begin{quoting}
 % \footnotesize
 %     \textit{I think the biggest barrier to entry is actually figuring out — what the heck you want to measure? And with AI safety, a lot of it is a bit wishy-washy. I feel a lot of the problems that people mention, especially sort of misalignment-related issues, have only been observed in these very laboratory settings.} --- \textbf{P6$_{S}$}
 % \end{quoting}

Human research is valued for constructing evidence of AI's impact, in terms of both benefits and harms. It can counter or validate normative assumptions embedded in AI design and evaluation (P4$_{S}$, P6$_{S}$, P12$_{T}$), and ground theory on socially complex topics like fairness, cognition, and human-AI collaboration (P5$_{T}$, P8$_{G}$, P10$_{T}$, P16$_{N}$). As P12$_{T}$ notes, \textit{``the really hard part is actually defining what is a bad output, and my work was assuming you have this defined"} --- a definition that cannot be assumed without evidence.

This need for human evidence is particularly acute for governance. P11$_{T}$ and P14$_{T}$ argue that policy should be supported by empirical evidence such as large-cohort A/B experiments, but this is not yet standard. P8$_{G}$ states that \textit{``the studies we need for AI policy are non-existent yet"}, and P7$_{G}$ echoes that \textit{``we have such little evidence on the effects on people."} In technical AI governance and alignment, interest centers on catastrophic and emergent harms. Incident case studies (e.g., the AI Incidents Database\footnote{\url{https://incidentdatabase.ai/}}) offer indicators of impending risk (P7$_{G}$), but human-validated evidence of harm at scale, or of effective mitigations, remains lacking.

 \begin{quoting}
 \footnotesize
     \textit{AI safety people are trying to be proactive and front-run against hypothetical harms, or harms that seem very likely to spring up but haven't concretely materialized yet. [...] There's a very fine, unclear line about when technology amplifies human agency or reduces it. A really important question that would be nice to solve is not necessarily a monolithic benchmark for this, but a collection of indicators that correspond with human disempowerment.} --- \textbf{P2$_{T}$}, on the need for human evidence for AIS problems.
 \end{quoting}

\begin{figure*}[t!]
    \centering
    \begin{subfigure}[t]{0.48\textwidth}
        \centering
        \includegraphics[height=2.1in]{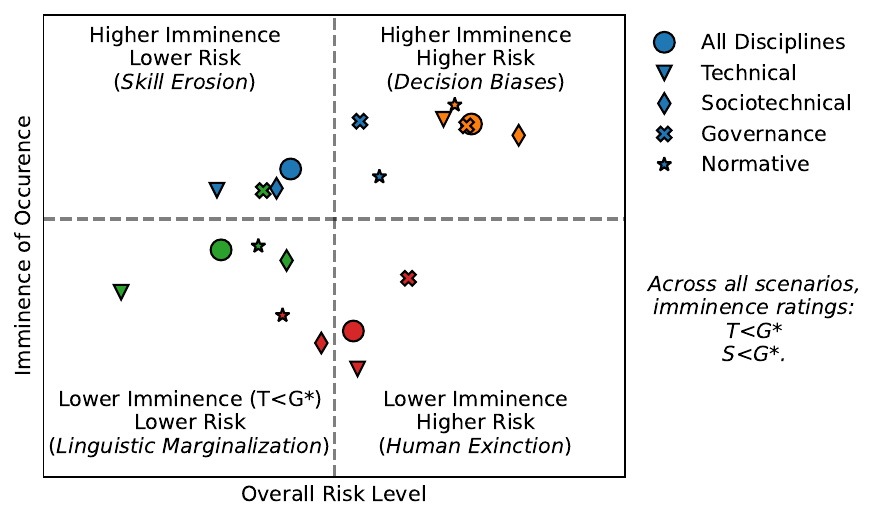}
\caption{Rating of the imminence and risk level (\textit{likelihood} $\times$ \textit{severity}) of four AISE risks. The average ratings (larger circles) agree with the a priori classification of the scenarios. The only inter-disciplinary differences is \textit{Governance} rating scenarios as more imminent. }
        \label{fig:risk_imminence}
    \end{subfigure}%
    \hfill
    \begin{subfigure}[t]{0.48\textwidth}
        \centering
        \includegraphics[height=2.1in]{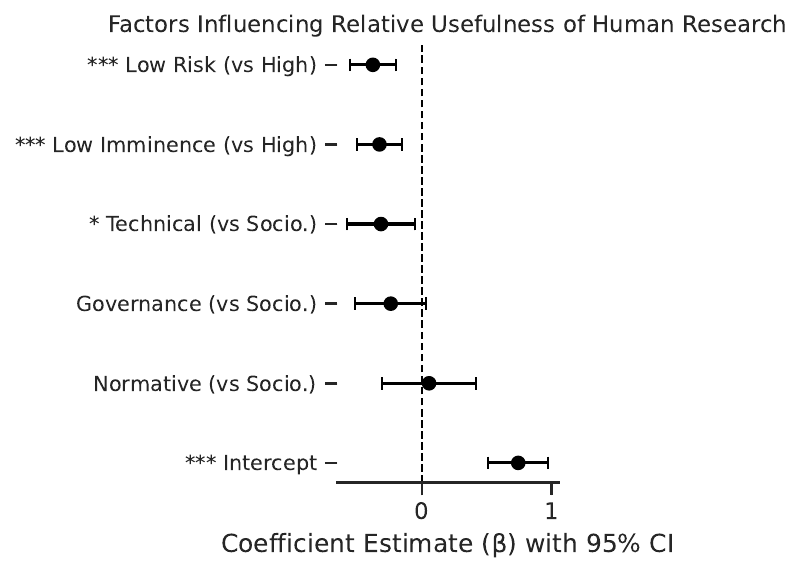}
        \caption{Mixed-effects estimates of factors influencing the relative perceived usefulness of human research methods. \textbf{Human methods are valued less when the scenarios have low risk and imminence, and by the \textit{Technical} discipline in comparison to \textit{Sociotechnical}.}}
        \label{fig:human_usefulness}
    \end{subfigure}
    \caption{Survey participants' ratings of AISE risk scenarios (left) and their preferences for human vs non-human research approaches for the scenarios (right). These ratings capture the general underlying beliefs, not discipline-specific practices.}
    \label{fig:scenarios}
\vspace{-0.5em}
\end{figure*}

The survey corroborates these interview findings. We measured both the prevalence of human research in participants' current practices and its perceived value if incorporated. Human empirical research was decomposed into four methodological dimensions: \textit{quantitative} vs. \textit{qualitative} paradigms, \textit{lab} vs. \textit{field} settings, \textit{experimental} vs. \textit{observational} designs, and \textit{representative} vs. \textit{specialized} sampling. For each, we computed a \textit{gap score} (aspiration $-$ practice) to quantify the discrepancy between what researchers do and what they would value doing. Results are shown in Figure \ref{fig:practice_aspirations}; full analysis details are in Appendix
% \footnotemark[2] 
\ref{app:survey_analyses}.

Consistent with disciplinary expectations, \textit{Sociotechnical} researchers practice and aspire to 
human research at significantly higher rates than \textit{Technical} (practice: $Z=4.19, p_{adj}<.001$; 
aspiration: $Z=2.69, p_{adj}=.007$) and \textit{Governance} (practice: $Z=2.67, p_{adj}=.007$; aspiration: $Z=2.79, p_{adj}=.005$), by Holm-corrected Dunn post-hoc tests following Kruskal-Wallis comparisons.\footnote{\textit{Normative} excluded in analyses due to small sample size.} Direct contrasts between \textit{Technical} and \textit{Governance} were largely non-significant, other than \textit{Governance} reporting more qualitative practice and \textit{Technical} showing stronger aspirational preference for experimental designs. Critically, across all disciplines and dimensions, gap scores were consistently positive, indicating that human research is valued beyond what is currently practiced.

At the research dimension level, Wilcoxon signed-rank tests revealed a single significant difference: researchers aspire to conduct \textit{field} studies more than \textit{lab} studies. This has two implications. First, no other dimension --- including the paradigmatically contentious quantitative/qualitative divide --- produced significant differences, suggesting that diverse forms of human research are broadly considered epistemically acceptable within AISE. Second, the field-over-lab preference signals a collective interest in \textit{ecological validity} over experimental control. In summary, there is clear interest for human research across disciplinary lines, but an aspiration-practice gap means it is not being conducted at a rate commensurate with the field's needs.

\subsubsection{\red{The Commensurability Gap.}}

% In this section, we contrast the extent to which disciplines perceive the relative importance of problems and solutions in AISE research differently from each other. In the survey, we ask experts to judge four AI-related safety scenarios and rate the usefulness of both human and non-human methods. Here, the participants are not asked to represent their own research or disciplines, so their answer captures what they \textit{generally} believe to be useful for the field. The results for scenario classification are in Figure \ref{fig:risk_imminence} and relative usefulness of human methods are in Figure \ref{fig:human_usefulness}. See Appendix \ref{app:survey_analyses} for more details of the statistical analysis.

We next examine whether and why disciplines differ in how they value human versus non-human methods. In the scenario-based survey section, participants rated the usefulness of both human and non-human methods for addressing four AISE risk scenarios varying on risk level (low vs. high) and imminence (future vs. current). Participants were not asked to represent their discipline, so these ratings reflect general beliefs about the field, albeit filtered through the lens of their epistemic training. Results are in Figure \ref{fig:scenarios} and the detailed analysis is in Appendix
% \footnotemark[2] 
\ref{app:survey_analyses}. The full description of scenarios and methods is provided in Appendix
% \footnotemark[2] 
\ref{app:survey}.

Participants agreed broadly on the classification of scenarios: \textit{Skill Erosion} and \textit{Decision Biases} were rated as more imminent, while \textit{Decision Biases} and \textit{Human Extinction} were rated as higher risk. The one disciplinary divergence was in imminence ratings: \textit{Governance} rated scenarios as more imminent than both \textit{Sociotechnical} 
($Z=2.62, p=.04$) and \textit{Technical} ($Z=3.25, p=.007$), consistent with governance researchers' 
orientation toward anticipating risks before they fully materialize. Otherwise, disciplines were in close alignment on perceived severity.

To test whether discipline predicted the relative value assigned to human  (e.g. `\textit{test intervention with users'}) versus non-human (e.g. \textit{`develop an automated benchmark'}) methods, we fit a mixed-effects model with random intercepts by participant, modelling the `human $-$ non-human' usefulness gap as a function of risk level, imminence, and discipline (reference: \textit{Sociotechnical}). The positive intercept ($\beta=0.74, p<.001$) confirms that \textit{Sociotechnical} researchers favoured human methods overall. This preference was attenuated for low-risk ($\beta=-0.38, p<.001$) and future scenarios ($\beta=-0.33, p<.001$), indicating that the perceived advantage of human methods diminishes when stakes are lower or harms more distal. \textbf{\textit{Technical} researchers showed a significantly smaller preference for human methods than \textit{Sociotechnical} ($\beta=-0.32, p=.017$)}; \textit{Governance} showed a similar trend that did not reach significance ($\beta=-0.24, p=.084$).

The interviews offer explanations for this observed undervaluation of human methods, stemming from a combination of legitimate methodological concerns and epistemic bias. On the legitimate side, construct and measurement validity were the foremost concerns, particularly for RCTs \cite{paskov2026rcts}. P8$_{G}$ critiques existing human-AI interaction RCTs as \textit{``really methodologically poorly designed and not controlled"}, and P6$_{S}$ echoes that some human studies \textit{``aren't credible because their methodology doesn't answer the [research question]."} P13$_{N}$ finds issues in how constructs like \textit{``trust in AI is measured and operationalized in empirical studies"}. Scalability is also a concern: frequently updated models cannot always be evaluated with human participants (P11$_{T}$), and large-population effects may be more tractable via simulation (P16$_{N}$).

Beyond legitimate concerns, discipline-specific biases also play a role. P6$_{S}$ and P8$_{G}$ both emphasized the importance of incorporating qualitative analysis to \textit{``explain idiosyncrasies and unique perspectives of individuals"} (P6$_{S}$) and \textit{``investigate how people use evidence and outputs of AI"} in complex, strategic tasks (P8$_{G}$). However, technically oriented researchers reported less familiarity with qualitative methods and sometimes held partial views of them --- perceiving qualitative evidence as perhaps \textit{``easier to manipulate than quantitative ones"} (P2$_{T}$). P12$_{T}$ notes that even in the relatively qualitative domain of red-teaming, reporting of \textit{how} attacks succeeded is often too thin to generalize or reproduce, reflecting underinvestment in the rigour of qualitative works. Human methods that map onto quantitative optimization are preferred. For example, `AI uplift' studies (RCTs of how AI scales human capabilities) are reported as the paradigm of choice for many (P2$_{T}$, P7$_{G}$). 
% Qualitative methods that could explain the mechanisms of harm receive less enthusiasm, despite their potential to address precisely the construct validity issues the field identifies. 

% We also identify a tension in the perceived value of methods. While many pointed out the positivist influence of machine learning over the epistemological norms of AI Safety, P6$_{S}$ and P8$_{G}$ both emphasized the importance of incorporating qualitative analysis to \textit{explain idiosyncrasies and unique perspectives of individuals} (P6$_{S}$) and \textit{investigate how people use evidence and outputs of AI} in complex, strategic tasks (P8$_{G}$). But researchers from more technical orientations admit to less familiarity with qualitative methods, and may hold certain biases like perceiving them as \textit{"easier to manipulate than quantitative ones"} (P2$_{T}$). Human methods that align closer with quantitative optimization are preferred, for example, "AI uplift" studies, which are AI Safety's RCTs that investigate how AI can scale up human abilities (P2$_{T}$, P7$_{G}$). There is also more interest in understanding how AI affects experts, who have greater capacity to enact good and harm from the boost from AI, rather than typical end users. In red-teaming studies, which are relatively qualitative methods for AIS, P12$_{T}$ points out that the reporting of \textit{how} red teaming succeeded often lacks details for reproducibility or generalizable value, pointing to how 

 \begin{quoting}
 \footnotesize
     \textit{I think [it needs to be] explained why qualitative studies are helpful. Yes, it's only 10 people, but you can still learn a lot from 10 people, as long as you ask the right questions and you do careful work. But because there's an obsession with scaling... and this is a barrier that is hard to overcome.} --- \textbf{P6$_{S}$}, on the misunderstanding of qualitative methods.
 \end{quoting}

\begin{figure*}
    \centering
    \includegraphics[width=\linewidth]{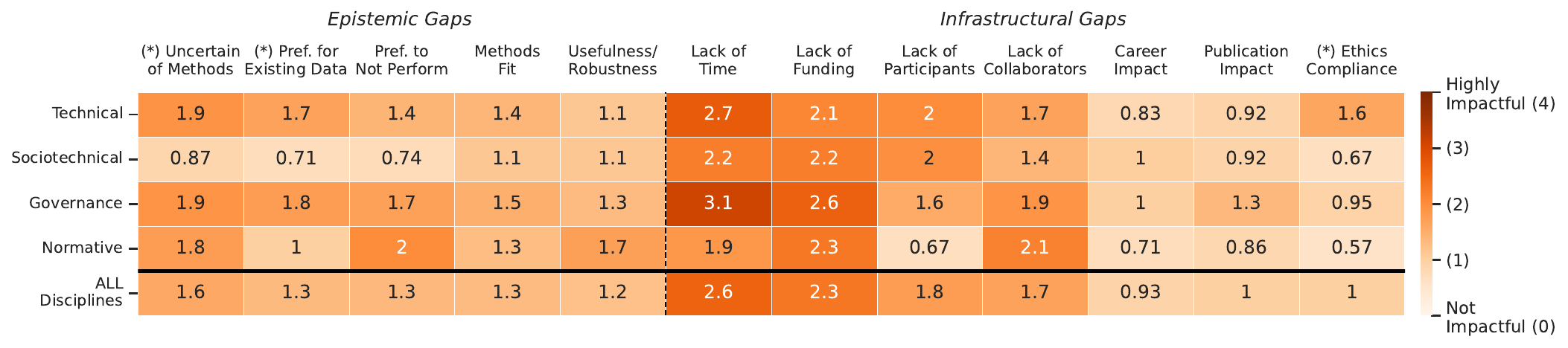}
    \caption{Ratings for the impacts of epistemic (RQ1) and tangible (RQ2) barriers to human research. Categories with a significant Kruskal-Wallis test of group differences are indicated with * ($p < .05$).}
    \label{fig:gaps}
\vspace{-0.5em}
\end{figure*}

The survey's methodological barriers section (see Figure \ref{fig:gaps}, which is combined with RQ2's analysis on tangible barriers) corroborates this: \textit{Sociotechnical} researchers reported significantly less impact from the categories of \textit{`uncertainty about methods'} and \textit{`preference for using existing datasets'} than other disciplines. Aesthetic and professional factors also play a role: some researchers stated to be deterred by the `messiness' of human data (P6$_{S}$, P12$_{T}$), while others cite a lack of prestige associated with empirical methods in their communities (P5$_{T}$). P2$_{T}$ suggests that HCI research may be more appealing to AIS researchers through \textit{``aesthetically changing"} the semantics and motivations used. Nevertheless, even technically oriented experts acknowledge the limits of purely computational approaches, as P9$_{S}$ reflects, \textit{``we try very hard to avoid humans, but then in some settings, we find that actually involving humans is the only way to answer the research questions."}

% The epistemic barriers section of the survey finds significant disciplinary differences in the impact from \textit{uncertainty of methods} and \textit{preference for using existing human data}, where \textit{Sociotechnical} reports significantly less impact. The inherent epistemic differences of the disciplines determine what types of evidence they see as rigorous, what methods they are familiar with through training and exposure in collaborations, and what aesthetic biases and preferences they absorb from their discipline. P2$_{T}$ suggests that research from fields like human-computer interactions may be more appealing to technical AI safety researchers through \textit{"aesthetically changing"} the semantics and motivations used. When commenting on others' inclination against human research, reasons given are that some are de-motivated by messiness (P6$_{S}$, P12$_{T}$), laziness (P6$_{S}$), and lack of prestige (P5$_{T}$). And despite having preferences for existing human datasets, P9$_{S}$ describes that although \textit{"we try very hard to avoid humans, but then in some settings, we find that actually involving humans is the only way to answer the research questions"}. 

% Intrinsic motivation differs, for some it is a necessary means to answer the research question (P1$_{S}$, P9$_{S}$), \textit{ we try very hard to avoid humans, but then in some settings, we find that actually involving humans is the only way to answer the research questions} (P9$_{S}$) 

\subsubsection{Challenges in Cross-Disciplinary Bridging.}
The preceding findings find agreement that human evidence matters, but persistent disciplinary siloes constrain its adoption. Here, we examine how those epistemic gaps manifest in collaboration and what conditions support bridging.

AISE (as well as the broader community of AI in general) is an evolving field. This is reflected in the 
expanding scope of technical AI conferences (P9$_{S}$), shifts in community composition (P6$_{S}$), and the fact 
that 77 of 93 survey participants identify with more than one research area. Yet in both cross-disciplinary engagement and collaboration frequency shown in Figure \ref{fig:participants}, \textit{Technical} researchers report the lowest, suggesting a stronger tendency toward disciplinary siloing in AIS. \textbf{\textit{Technical} collaborates significantly less than all other disciplines}(\textit{Sociotechnical}: $Z=-3.57, p_{adj}=.001$; \textit{Governance}: $Z=-3.45, p_{adj}=.001$) by Holm-corrected Dunn post-hoc tests following Kruskal-Wallis. As P13$_{N}$ emphasizes, \textit{``[AISE] problems are definitely too complex to have only one person in the room"}, but the field remains dominated by technical safety perspectives.

Tracking with prior research, the AIS/AIE divide is particularly salient \cite{Gyevnar2026-uq, roytburg2025mind}. Participants noted that while methods differ across these communities, their goals are aligned (P9$_{S}$) and \textit{``the methods are different, but the mental structure is the same"} (P13$_{N}$). The barrier is not disagreement about what matters, but access to knowledge and people. P13$_{N}$ described difficulty finding entry points into human-centred AI 
expertise, saying \textit{``connecting the experts in human-centered AI with people who are interested in human-centered AI is not easy."}
% ... are there textbooks on human-centered AI anywhere? That would be great for people like me."} 
At the same time, bridging must be bidirectional. HCI and social science researchers should also actively engage with AIS communities rather than waiting to be contacted (P6$_{S}$). Institutions that supports interdisciplinary exchange facilitates these collaborations (P1$_{S}$, P6$_{S}$, P16$_{N}$), but is not universally available (P12$_{T}$, P14$_{T}$, P17$_{S}$).

 \begin{quoting}
 \footnotesize
     \textit{It's important to acknowledge others' perspective… but we shouldn't expect fields to then just flatten their differences completely, right? Can we approach those differences in a structured way? Rather than immediately going into this reflex of, “I hate this”, or “I love this”?} --- \textbf{P6$_{S}$}, on constructively bridging between different epistemologies. 
 \end{quoting}

Experts appreciated cross-disciplinary venues like IASEAI, but felt that the community still reflected AIS over AIE, as \textit{``the safety groups have money and presence"} (P15$_{G}$), and the opportunities for bridging were superficial and \textit{``didn't quite work this year"} (P6$_{S}$). Top-down unification efforts carry risks of epistemic collapse and academic gatekeeping, and P6$_{S}$ observes that \textit{``AI safety has not yet matured into a field where human studies are accepted as a necessary component of evaluation."} This creates a dual risk. If human research is excluded from AISE's epistemic boundaries, the evidence base for making AI systems safe will be fundamentally incomplete. But if human research adopted as a performative checklist item rather than a rigorous contribution, the result is what we term \textit{human-washing}: the superficial inclusion of human subjects that creates the appearance of validity. Both outcomes diminish the field's capacity to generate credible evidence of harm and advance the objective of making AI safe and ethical.

\subsection{RQ2: Barriers to Human Research: Resources, Infrastructure, and Sector Boundaries }

Having examined how AISE experts value human research and why a gap persists between aspiration and practice, we now turn to the tangible barriers that prevent researchers from executing human studies. We identify three levels of constraint: resource-related, infrastructural, and sector-level dynamics, reflecting prior findings on barriers of human research \cite{agapie2022using}. Survey ratings on the degree to which each barrier was felt are shown in Figure \ref{fig:gaps} and interview accounts provide explanatory depth.

% The previous section examines how AISE experts consider the value of human research, and touches on the gap between aspirations and practice. In this section, we narrow in on our participants' lived experiences as AISE researchers, exploring the \textit{tangible barriers} they face in executing human research. We identify and describe three levels of limitations that contribute to the human empirical evidence gap, outside of epistemic suitability: resources, infrastructure, and sectors. The themes are supported with survey ratings on the degree of impact participants faced from barriers, shown in Figure \ref{fig:gaps}. Note that the set of barriers we asked in the survey does not cleanly overlap with the ones discussed in the interviews, as the interviewees covered additional topics relevant to them. In the survey, people were allowed to select \textit{Not Applicable} if the barrier did not apply to them. One limitation is that the questions were not contextualized to specific projects, so we cannot retroactively disentangle to what extent the limitations were experienced \textit{in the process} of doing research, or \textit{prevented} the execution of the research altogether. 

\subsubsection{Resource Limitations.}
The most direct constraints concern the material resources available to researchers: time, funding, and high-quality participants. No significant differences across disciplines were found for any resource barrier in the survey, suggesting these constraints are felt broadly. Averaged across all respondents, \textit{lack of time} was ranked most impactful, followed by \textit{funding}, then \textit{participant access}. These barriers often compound each other, such as long recruitment periods consuming time that delays other work, and waiting on funding decisions stalls momentum. Ethics approval adds further administrative delay, which P4$_{S}$ notes can disrupt the continuity of a research project.

% The most limiting and direct factor is the project resources that researchers have access to. This includes funding, time, and high-quality participants. There were no significant differences across disciplines in any of the resource barriers asked in the survey. A lack of time was ranked as the most impactful item averaged across all respondents, followed by funding, then participants. Interactions between the barriers, like waiting for funding or recruiting for prolonged periods, amplified the hindurances felt by the researchers. Time, in particular, may be the most apprehension-inducing factor because it is seen as a waste of productivity that can be better spent on other career-advancing activities. P4$_{S}$ also notes that ethics approvals pose significant administrative delays and impact the momentum of sustaining research. 

 \begin{quoting}
 \footnotesize
     \textit{I had a discussion with [a colleague] who's doing large-scale human subjects research, and he just wanted to go back and do a [computational] benchmark... you can spend a few months and get a good result, and then it's done.} --- \textbf{P9$_{S}$}, on the substantial time required for human research.
 \end{quoting} 
 
% While time is often weighed internally by the researcher, funding poses a hard external constraint. P10$_{T}$ was blocked from conducting a neuroscience based human-AI interactions study after losing out on a sizeable grant. The estimated costs of 300k USD was prohibitive against continuing the research, despite having a developed proposal already. In contrast, P9$_{S}$ estimates the cost of computational experiments to be in the magnitude of \textit{"less than a thousand dollars per paper"}. Funding issues were variable, however, as P1$_{S}$ and P4$_{S}$ reported receiving adequate support from their supervisors. P6$_{S}$, who participates in grant reviews, finds promising movement towards the inclusion of sociotechnical research topics in AI Safety grants. 

While time is often weighed internally by the researcher, funding poses a hard external constraint. P10$_{T}$ was unable to conduct a neuroscience-based human-AI interaction study due to losing a major grant. The estimated cost of 300k USD was prohibitive despite a fully developed proposal. This sharply contrasts with the cost of computational work, which P9$_{S}$ estimates at \textit{``less than a thousand dollars per paper."}  However, funding conditions vary, as P1$_{S}$ and P4$_{S}$ reported adequate support from their labs. P6$_{S}$, who participates in grant reviews, observed promising movement towards including sociotechnical research in AISE funding.

Participant access introduces both logistical difficulty and quality trade-offs. P1$_{S}$ underwent multiple recruitment rounds after encountering scam respondents in a study requiring highly specific participant qualifications, reflecting the increasing issues in fraudulent participants observed in HCI \cite{panicker2024understanding}. P4$_{S}$, P11$_{T}$, and P13$_{N}$, who work in health-specific contexts, describe difficulty accessing medical specialists and clinical populations. Field studies present additional complexity, as P15$_{G}$ notes that in these contexts, \textit{``you're really dependent on partnership building, partnership quality, and it might be 10 times as much time investment for a smaller scale."} Recruiting non-specialist populations such as crowdworkers or undergraduate students is more attainable (P6$_{S}$), but is perceived to reduce the relevance and validity of the findings (P2$_{T}$).

% Participant recruitment also contributes to both the time delay and logistical difficulties. P1$_{S}$ describes undergoing multiple rounds of recruitment due to encountering scammers in a study that is further complicated by the specificity of the qualifications participants must meet. Similarly, P4$_{S}$, P11$_{T}$, and P13$_{N}$ work in specialized health contexts and describe difficulty attaining access to participants and medical specialists. P15$_{G}$ encountered setups in field studies \textit{"where you're really dependent on partnership building, partnership quality, and it might be 10 times as much time investment for a smaller scale"}.  Running studies with non-experts, like crowdworkers or undergraduate students, is much more attainable (P6$_{S}$) but is seen to dilute the value of the work (P2$_{T}$). 

% Lastly, P5$_{T}$ and P9$_{S}$ describe an internal limitation in the lack of personal bandwidth to perform face-to-face human interviews, with P5$_{T}$ describing that \textit{"it's stressful, and I have a lot of social anxiety"}, and P9$_{S}$ describing their interviewing approach as akin to doing a \textit{"podcast interview"} without formal training. 

\subsubsection{Infrastructural Limitations.}
Beyond individual resources, barriers are embedded in the \textit{infrastructure} surrounding researchers --- the organizational arrangements, mentorship structures, and professional incentives that shape what research is possible and viable to sustain careers \cite{lee2006human}. These constraints fall heaviest on junior researchers, who depend most on their immediate environment for access and legitimacy.

% We use \textit{infrastructrual}, as defined by \citeauthor{lee2006human}, in reference to the \textit{"arrangement of organizations and actors that must be brought into alignment in order for work to be accomplished".} This category concerns the issues within a researcher's broader organizational structure, including opportunities to receive mentorship and collaboration, that impede their exposure to or progress in human research. Without adequate support structures, junior researchers are at risk for the greatest impacts to their career development. 

Mentorship is a significant mechanism of both restriction and access. \textit{Normative} participants rated \textit{'lack of collaborators'} as one of their highest impact barriers, and the highest of any discipline. More broadly, interviews captured accounts of advisors actively discouraging human research: P9$_{S}$'s mentors were opposed to working with human subjects; P5$_{T}$ and P16$_{N}$'s institutions and leadership did not recognize the value of empirical methods; and P17$_{S}$'s sociotechnical interests were directly in tension with their advisor's focus on AI. In the absence of formal interdisciplinary structures, junior researchers are left to build networks informally, either through cold outreach (P17$_{S}$) or student communities (P2$_{T}$, P16$_{N}$, P17$_{S}$).

% Senior mentorship played a significant role in influencing and restricting access to human research. \textit{Normative} participants in the survey  rated \textit{lack of collaborators} the highest among the disciplines, and second highest across all barriers. The interviews capture experiences from P9$_{S}$, whose mentors advocated against working with human subjects; from P5$_{T}$ and P16$_{N}$, whose broader research organizations and leadership do not recognize the value of empirical methods; and from P17$_{S}$, whose interest in sociotechnical impacts conflicts with their advisor's traditional approach to AI research. In absence of a formal collaboration structure, P17$_{S}$ resorted to reaching out to mentors themselves, where student clubs are cited as a way to meet like-minded people (P2$_{T}$, P16$_{N}$, P17$_{S}$).
 \begin{quoting}
 \footnotesize
     \textit{My advisor is a standard computer science, technical person who just wants to improve on the benchmarks and achieve AGI. I don't think I've been able to get through to him, so my goal with my PhD is to just shoehorn in the things that I think are important, while on the surface making it still look like an AI research PhD.} --- \textbf{P17$_{S}$}, on navigating tensions between their interests and their advisor's. 
 \end{quoting}
 
% We also include external pressures from limitations to future career opportunities. While the survey's lower ratings for \textit{career impact} and \textit{publication impact} indicate that these factors are not felt broadly across the board, it can be highly disadvantageous for select individuals. 
% In particular, P5$_{T}$ recounts struggled in establishing grounds in collaborations as a researcher bridging between an HCI environment and a theoretical one. Meaningful cross-discipline work is difficult due to many underlying differences in the publishing standards of the fields. In addition, they encountered gendered struggles around establishing the legitimacy within a highly theoretical research group, which deters them from pursuing empirical work with humans. 

Career and publication pressures, while not highly rated in the survey overall, can be acutely constraining for specific individuals. P5$_{T}$ describes difficulty establishing credibility at intersection of HCI and theoretical AI research, compounded by the incompatible publication norms of those fields. They also recount gendered barriers around establishing legitimacy within a highly theoretical research group, which deters them from pursuing human research. This points to a broader pattern in AISE that privileges formal, quantitative, and technical contributions. These implied epistemic values reflect whose research is seen as legitimate. 

 \begin{quoting}
 \footnotesize
     \textit{I want to be accepted as a theoretical researcher. Because if I do human studies, people will think I am just here because I'm a woman, and I'm doing `woman' research.} --- \textbf{P5$_{T}$}, on gendered biases faced in theoretical research.
 \end{quoting}

\subsubsection{Sector-Level Constraints.}

Barriers to human research also vary by sector, and opportunities for mitigation lie partly in cross-sector collaboration. In non-profits and industry, the absence of formal ethics review infrastructure impedes engagement with human subjects, particularly in high-risk research contexts (P8$_{G}$, P11$_{T}$, P14$_{T}$) \cite{metcalf2016human}. P8$_{G}$, a policy manager at a non-profit research organization, addresses this by funding academic student fellowships, reflecting a broader view that human research is best conducted in academic institutions, which provide ethical oversight and research neutrality.

% For the last theme, we describe the interplay between sectors (academia, non-profit, industry, and government), where each faces their own limitations, but can benefit from stronger cross-sector collaborations.  For example, in non-profits and industry, challenges in meeting ethical standards while lacking a formal ethical review board hold them back from engaging with human subjects altogether, especially in high-risk contexts (P8$_{G}$, P11$_{T}$, P14$_{T}$). P8$_{G}$, a manager in a non-profit research institute, directs funds towards student fellowships in engagement with academia. Human research is believed to be best performed in academia, or with strong involvement of academics as neutral parties. 

  \begin{quoting}
 \footnotesize
     \textit{A lot of evidence from the labs is coming from companies that have business incentives to give impression that the models they are producing are extremely robust. Everyone thinks, ``the AI labs have the best scientists, so they know what they're talking about", which I think is very problematic.} --- \textbf{P8$_{G}$}, on ethical issues of industry research.
 \end{quoting}

The involvement of industry in AISE research is a source of both problematic implications and opportunity \cite{abdalla2021grey}. Participants raised concerns about the influence of unregulated, commercially incentivized actors on the research landscape (P2$_{T}$, P11$_{T}$), and about the rigour and neutrality of industry-produced research (P8$_{G}$, P13$_{N}$, P15$_{G}$). For example, Facebook's emotional contagion study \cite{kramer2014experimental} is mentioned as a reference for ethical failure (P9$_{S}$, P15$_{G}$). At the same time, industry's access to large-scale deployment data and high-quality research talent cannot be overstated. P15$_{G}$ argues that, \textit{``if [companies] doing risky research, or they want their research to be rigorous and not just internal A-B testing, I think they should be more academic."} Interesting, P2$_{T}$ adds that this credibility tension is not one-directional, saying \textit{``that there's probably some prejudice from the AI safety community's perspective, where [AI Safety thinks] a lot of the good research hasn't come from traditional academia."}

Strengthening cross-sector collaboration is broadly viewed as necessary given the pace of AI development (P15$_{G}$). Governance and policy decisions are increasingly reliant on evidence that does not yet exist at the required scale, but in turn, these decisions will play a substantial role in controlling the speed and veracity of AI development. P3$_{G}$ further focuses on international politics, drawing comparisons to prior large-scale collaborations between countries in times of heightened tension. Ultimately, the timeline for multi-disciplinary, cross-sector, and international collaboration to prevent both existential and cumulative risks of AI is an urgent, epistemic priority.

% Ultimately, having more sectors involved --- especially given the rapid proliferation of AI --- is a positive impact on actionability (P15$_{G}$). Across previous sections, we described how governance is deeply reliant on evidence of harm and general guidelines, like the International AI Safety Report (P7$_{G}$) and the United Nation's seven recommendations for AI (P3$_{G}$), etc, act as key references. Governance and policy decisions will in turn play a substantial role in the controlling the speed and veracity of AI development. P3$_{G}$ further focuses on international boundaries, drawing comparison to prior large-scale collaborations between countries at times of heightened tension, but urging the timeline for collaboration to prevent existential risks of AI. 
%   \begin{quoting}
%  \footnotesize
%      \textit{The AI community is in its infancy with that, because they're all just still figuring out. It took 7 years to negotiate the IAEA at the height of the Cold War and it took 20 years to negotiate the ITER agreement. Normally, these things take a long time. } --- \textbf{P3$_{G}$}
%  \end{quoting}

\section{Discussion}
%\subsection{Key Findings}
Our results show that experts across disciplines agree that empirical research with human participants offers meaningful evidence to AISE --- and yet, despite epistemic fit, it remains \red{undervalued}. Although experts demonstrated interest for various methods of human research, many highlighted a predominant positivist bias that favours more technical methodologies and metrics, limiting the field's ability to answer research questions that require interpretivist and qualitative empirical approaches. Additionally, practical barriers and individual researcher biases further reinforce dominant research priorities. As the AISE community continues to iteratively define its identity, top-down efforts in constructing the epistemic boundaries of what methods and whose voices get to be highlighted must be approached critically. We synthesize recommendations and implications for bridging the epistemic gaps (RQ1) and tangible barriers (RQ2).

\subsection{Epistemic Implications for AISE Researchers}
\subsubsection{Field-Level Epistemic Boundaries.} 
Multidisciplinary fields such as human-computer interaction, public health and environmental sciences have long struggled to bridge epistemic divides, as integrating different values, methods and ways of work takes intentional efforts throughout the entire life-cycle of a research project \cite{talbi2025reflections}. The framework of \textit{epistemological pluralism} \cite{miller2008epistemological} proposes reconciling disciplinary differences by recognizing them as complementary instead of competing perspectives. AISE faces a similar imperative. Our results demonstrate \textit{Technical} researchers systematically undervalue human methods relative to their \textit{Sociotechnical} counterparts, suggesting that the epistemic divide in AISE is actively shaping what evidence gets legitimized. Left unaddressed, this tension may compound the risk of ideological homogeneity in AISE community-building \cite{Ahmed2023-cr, Dahlgren-Lindstrom2025-ah}. Thus, we pose that AISE should welcome the challenge of incorporating different `ways of knowing' to expand potential avenues to identify, measure and mitigate AI harm \cite{Gyevnar2025-uw}. %Achieving this requires purposeful leadership and explicitly inclusive epistemic boundaries from the organizations that shape the field.

\subsubsection{Methodological Literacy for Researchers.} Beyond field-level recommendations, improving the presence of human research in AISE also requires change at the individual researcher level, particularly among \textit{Technical} researchers, who report the lowest cross-disciplinary engagement. We do not argue that all researchers should conduct human studies -- after all, disciplinary divides can provide constructive value \cite{Gyevnar2026-uq} -- but that practitioners should understand the established methodological literature of adjacent fields \cite{talbi2025reflections}. \red{Researchers should explicitly recognize incommensurability as a barrier to accepting different ways of generating evidence for AISE \cite{politi2017specialisation}.} The case of AI uplift studies illustrates both the promise and the limits of current practice. Despite involving human participants, uplift studies largely reproduce the positivist paradigm dominant in machine learning: controlled lab conditions, aggregate effect sizes, and quantitative outcome measures. Practitioners have themselves acknowledged core challenges around task realism in human uplift studies leveraging RCTs \cite{paskov2026rcts}, yet the body of HCI and social science research that has theorized and addressed exactly these ecological validity problems is rarely engaged. This insularity reflects a broader pattern: AISE's dominant positivist orientation privileges measuring the likelihood of harm while overlooking the interpretivist methods needed to understand the mechanisms through which harm unfolds and how it is experienced in practice. 
% Cross-disciplinary literacy is a bidirectional obligation; researchers from human-centered disciplines should equally engage with the technical AI safety communities for meaningful collaboration and resource sharing. Mutual understanding of each others' disciplines strengths and shortcomings is paramount to establish collaboration that can unlock forward-thinking approaches and generate substantiated contributions.
 
\subsubsection{Prevent ``Human-Washing".} While encouraging broader adoption of empirical human research methods in AISE, we must equally caution against superficial or performative participant engagement, as we termed \textit{human-washing}. Past work has flagged similar patterns in machine learning research, like \textit{diversity washing} \cite{whitney2024real} and \textit{participation washing} \cite{sloane2022participation} that result in harmful misrepresentation and exploitation instead of inclusion. \textit{Human-washing} poses the same risk, where the inclusion of human participants is enacted without the necessary methodological rigour, creating the appearance of empirical grounding while undermining it. Our interview participants note the concern that some existing human studies in AISE are insufficiently designed to answer the research questions they claim to address. 
% \red{currently there is evidence gap, but there is also commensurability gap.} 
As the field's epistemic boundaries evolve, and as venues like IASEAI expand their scope, there is a potential risk that human research becomes a checklist requirement. Mandating human research without cultivating the literacy and infrastructure to do it well would create downstream issues with validating evidence legitimacy.

\subsection{Implications for Resources and Infrastructures}
%To address the practical and systemic barriers to performing human research, we call upon institutions in academia, industry and non-profit sectors to place the incentives to foster interdisciplinary research collaboration. 

\subsubsection{Overcoming Resource Challenges}
The most direct lever available to institutions is funding, and the cost asymmetry between technical and human research is stark: a computational paper can cost under a thousand dollars, while a rigorous human-AI interaction study can run into the hundreds of thousands. Funding bodies, like government agencies, philanthropic foundations, and industry research programs, should explicitly account for the higher cost structures of human research in their grant mechanisms to acknowledge the imbalance and incentivize appropriately. Equally important is the framing and terminology used to describe funding opportunities, as opting for technical topics implicitly signals what methods are valued, discouraging human-centered proposals before they are written. This pattern can be observed in past grants that primarily targeted technical AI safety, with a focus on AI alignment: OpenAI's Superalignment Fast Grant\footnote{\url{https://openai.com/index/superalignment-fast-grants/}}, the AI Security Institute's Alignment Project\footnote{\url{https://alignmentproject.aisi.gov.uk/research-agenda}}, and so forth. Beyond funding opportunities, technical researchers without prior human subjects training would benefit from accessible resources for navigating ethics compliance so that Institutional Review Board (IRB) processes do not pose a significant barrier to entry.

%\red{From the funding perspective, grants have primarily targeted technical AI Safety research, with a focus on AI Alignment: OpenAI's Superalignment Fast Grant\footnote{\url{https://openai.com/index/superalignment-fast-grants/}}, the AI Security Institute's Alignment Project\footnote{\url{https://alignmentproject.aisi.gov.uk/research-agenda}}, and CIFAR's AI Safety Catalyst Grants\footnote{\url{https://cifar.ca/cifarnews/2025/06/04/cifar-announces-first-ai-safety-catalyst-grants-under-new-national-program/}}.}
% Lack of research resources are identified by survey participants as the most limiting factors. 
% For funding sources like government grants, industry research grants, and fellowship programs, they should be more inclusive and sensitive to the requirements of human research as opposed to listing only AI alignment topics. 
% On ethics, technical researchers should receive adequate training and resources to learn how to apply for IRB. And institutions can try to speed up the review process to make it not inhibitive to the timelines demanded by AI Safety.
% what recommendations can we make about participants?

\subsubsection{Mentorship and Training.}
Mentorship functions as a significant mechanism of methodological gatekeeping in AISE, with junior researchers who pursue human studies frequently facing discouragement from supervisors, limited access to relevant collaborators, and in some cases the need to obscure their methodological interests entirely. Departments and graduate programs should formalize pathways for interdisciplinary mentorship so that students are not wholly dependent on a single advisor's agenda and skill set \cite{jacobs2009interdisciplinarity}. Mitigation strategies such as co-supervision arrangements, visiting researcher programs, and funded rotations are examples to facilitate cross-pollination and to build stronger collaboration networks. Fellowship programs represent a parallel opportunity: existing AIS fellowships at non-profits and industry organizations have demonstrated an effective model for training junior researchers, but currently reflect the field's methodological insularity. Expanding these programs to include human-research tracks, or funding parallel structures in sociotechnical and governance areas, would both develop empirically literate researchers and signal that human methods are valued in high-prestige settings.
% As described by junior researchers, mentorship plays a large role in what research methods you get exposed to and trained in. In the case of strong tensions between the researcher and their immediate supervisor or department, juniors should have opportunities to find collaborators or mentors in the relevant areas. For example, funded research fellowship opportunities to work on AIS research exists in many AIS non-profits and companies (including the likes of the MATS program and OpenAI's fellowship). However, these programs currently do not embrace human methods. So more opportunities to do this would be beneficial. Also in academia departments, strong cross disciplinary department involvement can introduce students to people that they align with. 

\subsubsection{Cross-Sector Collaboration.} Our participants suggested a collaboration model structured around integrated academic-industry partnerships, rather than isolated parallel efforts. Additionally, non-profit organizations are well-positioned to act as neutral intermediaries, insulating research directives from commercial influence while preserving access to industry resources and facilitating data-sharing arrangements that could give academic researchers access to deployment-scale human evidence under independent ethics oversight. At the policy level, governance bodies should treat the production of human evidence as an active priority, funding the infrastructure that makes it possible rather than waiting for it to emerge organically. International coordination will be necessary for evidence about cross-jurisdictional harms, and while precedents from large-scale scientific collaborations suggest this is achievable, the pace of AI development means the window for establishing this infrastructure is already compressed. 

%Industry's advantages — deployment-scale data, frontier model access, and substantial research talent — are not replicable in academia, yet direct commercial incentives compromise the neutrality of safety claims produced entirely within labs. The model suggested by participants, and supported by precedent in other high-stakes research domains, is structured academic-industry partnership rather than either insularity or uncritical integration. Non-profit organizations are well-positioned to act as neutral intermediaries, insulating research from commercial influence while preserving access to industry resources, and pre-competitive data-sharing arrangements, analogous to those used in pharmaceutical research, could give academic researchers access to deployment-scale human evidence under independent ethics oversight. At the policy level, governance bodies should treat the production of human evidence as an active priority, funding the infrastructure that makes it possible rather than waiting for it to emerge organically. International coordination will be necessary for evidence about cross-jurisdictional harms, and while precedents from large-scale scientific collaborations suggest this is achievable, the pace of AI development means the window for establishing this infrastructure is already compressed. 

\subsubsection{Influence and Power.} As custodians of knowledge production, institutions in AISE exert substantial influence over which research agendas are legitimized and prioritized, with consequences that extend beyond the field into public discourse and policymaking. Industry research from the tech giants Anthropic, Google and OpenAI have, until very recently, concentrated efforts on the evaluation of AI abilities and behaviours \cite{achiam2023gpt, kalai2025language} over how real users engage with AI \cite{aiskillformation2026, phang2025investigating}. Notably, industry papers tend to receive more citations than those authored by researchers in academia \cite{strauss2025real}, further demonstrating their contributions receive more attention and traction. This trend shapes what gaps are considered worth filling, and what methods are considered rigorous. Potential mitigations include conflict-of-interest disclosure requirements at journals and conferences, and citation audits in systematic reviews to surface whose work is or is not being built upon. 

Furthermore, the concentration of prominent voices in AISE compounds this dynamic. Researchers from elite universities and major laboratories often share epistemological assumptions favoring technical and quantitative approaches that collectively marginalize the human-centered, qualitative and participatory methods. When the figureheads of AISE do not practice or value human-centered research methods, it is difficult for such methods to gain traction regardless of their epistemic merit. As a remedy expert panels, advisory boards, and parliamentary testimony should mandate disciplinary and methodological diversity, explicitly recognizing that lived experience of AI harm is a form of evidence that technically-oriented research will not reveal.

\subsection{Limitations}
Our findings should be interpreted with the following limitations. Recruitment was targetted to IASEAI attendees and the researchers' existing professional networks rather than a systematic sampling of the entire AISE community. The survey was also skewed toward WEIRD contexts, and industry researchers were underrepresented relative to their influence. Voluntary interview participants were likely already predisposed toward human research; this means those who are most skeptical of human methods might be underrepresented in our qualitative findings. \red{However, since we undersample researchers who are likely to be resistant to human research, the effect sizes observed in the quantiative analyses are optimistically stronger in reality}. While our four discipline categories provided a useful organizing structure, they were not intended to capture the full complexity of the researchers' identities. \red{For example, we did not collect detailed political or critical orientations of the participants, which can influence the biases in self-rated perceptions.} \red{Our survey questions were generated as an one-off data collection opportunity and are not meant to represent generalizable instrument scales.} Finally, because the interviews were intentionally open-ended, the themes that emerged did not always map onto the survey constructs, which limits direct comparisons between data sources. 

% \begin{itemize}
%     \item recruitment
%     \item survey design 
% \end{itemize}

%ultimately, human methods may still considered to be outside the epistemological boundaries of AISE (especially AIS) by a large contingent of researchers. we expect that as bridging between the disciplines to incrementally emerge, the role of human research may settle into an "accepted" or "normalized" regions. AISE will have to content with how to integrate the positivist paradismgs from quant and ML with the interpretivist paradigms from qual and social science. However, we pose that practices of using human research to unfaithfully demonstrate validity of results, or human-washing, should be critically examined and avoided. There is obviously a strong epistemic fit for human research in AISE to construct evidence, both upstream and downstream of theoretical, technical, and policy work. We expect that the evolution of the field will depend on the leadership / shared venues that dictate the inclusion of human evidence, human researchers, and human paradigms 

\section{Conclusion}
Navigating epistemic tensions is a natural step in the development of an emergent scientific field, and AISE is no exception. Nonetheless, past work has identified the predominance of technical AI Safety research paradigms in AISE literature. We contribute to this discussion by examining the lived experiences of the research practitioners in AISE via expert survey and interviews. Our findings reveal overall agreement that empirical human research adds value but is underutilized \red{and undervalued}. Critically, experts from \textit{Technical} backgrounds rate human methods as \textit{less useful}, and tend to \textit{collaborate less} across disciplines.
% This pattern is compounded by AISE's existing literature and funding incentives also favouring more technical lenses. 
This imbalance risks reducing the AISE research agenda to the point of epistemic hegemony, implicitly delegitimizing other approaches such as human-centred methods. We pose that explicitly inclusive, yet not performative, epistemic boundaries are paramount to the recognition of  evidence from human research paradigms to holistically address AI-related harms.

\section{Acknowledgments}
We would like to sincerely thank our interview and survey participants for making time to engage with our study. JYB is supported by the Vanier Canada Graduate Scholarship (FRN 198876), administered through the Natural Sciences and Engineering Research Council of
Canada (NSERC).

\section{Position Statement}
The academic and research backgrounds of the research team heavily informed the research questions and analytical methods used in this paper. All authors are academics from Computer Science and Information Sciences, with a focus on human-computer interactions, responsible AI, and systems safety. Everyone has experience with conducting human empirical research, ranging from high-powered controlled experiments to ethnographic studies. The first and last authors come from quantitative-leaning backgrounds in HCI and computational social science, and other authors have expertise in qualitative and critical methods. We therefore chose a mixed-methods approach that combines surveys and interviews. Three of the authors, including the first author, also engage with technical AI safety and have direct experience with synthesizing from multidisciplinary research. 

\bibliography{aaai2026}

% \clearpage

\clearpage

\appendix
\setcounter{secnumdepth}{2}

\section{Survey Questions}
\label{app:survey}

\subsection{Background and Experience}

\paragraph{Primary Research Area.}
What is the broad area of AI research that your work primarily aligns
with? 
\begin{itemize}
  \item Technical (model training/evaluation, AI safety/alignment,
    interpretability, proofs)
  \item Normative (philosophy, ethics theory, conceptual work)
  \item Sociotechnical (psychology, HCI, STS, empirical AI
    ethics/fairness, social science)
  \item Governance (policy, law, economics)
\end{itemize}

\paragraph{Secondary Research Area.}
If applicable, is there a secondary area of Safe \& Ethical AI research that your work aligns with?
\begin{itemize}
  \item Technical (model training/evaluation, AI safety/alignment,
    interpretability, proofs)
  \item Normative (philosophy, ethics theory, conceptual work)
  \item Sociotechnical (psychology, HCI, STS, empirical AI
    ethics/fairness, social science)
  \item Governance (policy, law, economics)
  \item N/A
  \item Other: 
\end{itemize}

\paragraph{Collaboration.}
How much do you collaborate with each area of Safe \& Ethical AI research?
\medskip\noindent \\
\underline{Response scale}: 1 (\textit{Never}) ---
2 --- 3 --- 4 --- 5  (\textit{Always})
\begin{enumerate}
  \item Technical (model training/evaluation, AI safety/alignment,
    interpretability, proofs)
  \item Normative (philosophy, ethics theory, conceptual work)
  \item Sociotechnical (psychology, HCI, STS, empirical AI
    ethics/fairness, social science)
  \item Governance (policy, law, economics)
\end{enumerate}

\paragraph{Specific Discipline.}
\noindent Please write your specific discipline
(e.g., model evaluation, critical computing, policy\ldots).

\paragraph{Education.}
What is your highest level of education?
\begin{itemize}
  \item Bachelor's (current or completed)
  \item Professional Master's (current or completed)
  \item Research Master's (current or completed)
  \item PhD (current)
  \item PhD (completed)
\end{itemize}

\paragraph{Sector.}
What sector do you work in?
\begin{itemize}
  \item Industry (Private Sector)
  \item Government (Public Sector)
  \item Non-Profit
  \item Academia (Student)
  \item Academia (Faculty/Researcher)
\end{itemize}

\paragraph{Role Type.}
What category does your role fall under?
\begin{itemize}
  \item Researcher
  \item Engineer
  \item Data Scientist / Analysis 
  \item Policy Analyst
  \item Manager
  \item Director
  \item Other:
\end{itemize}

\paragraph{Country.}
In which country do you primarily work? 

\paragraph{Research Involvement.}
To what degree are you engaged in research\footnote{Research is not restricted to traditional academic publishing. It encompasses alternative formats, like data collection for policy and research dissemination via in-depth preprints and blog posts.} activities in your current role?  
\begin{itemize}
  \item Primarily research-focused (e.g. research is full-time) 
  \item Substantial involvement (e.g. research is part-time) 
  \item Moderate involvement (e.g. research is occasional) 
  \item Little involvement (e.g. keeps up with literature but does not actively do research)   \item Manager
  \item Very little involvement 
\end{itemize}

\paragraph{Experience.}
How many years of experience do you have conducting research on AI or
AI-related topics? \hfill \textit{[0--10+]}

\subsection{Practice and Aspirations}

\paragraph{Practice.}
How often do you perform the following types of human empirical research in your Safe \& Ethical AI work?
\medskip\noindent \\
\underline{Response scale}: 1 (\textit{Never}) ---
2 --- 3 --- 4 --- 5  (\textit{Always})
\begin{itemize}
  \item Human empirical research (overall) 
  \item Quantitative research (e.g., behavioural and survey data) 
  \item Qualitative research (e.g., interview data) 
  \item Lab studies (controlled setting) 
  \item Field studies (real-world setting) 
  \item Experimental studies (comparing multiple conditions) 
  \item Observational studies (examining existing phenomenon) 
  \item Representative studies (generalizable findings) 
  \item Targeted community studies (context-specific 
\end{itemize}

\paragraph{Aspiration.}
To what extent would you incorporate the following human empirical research methods in your Safe \& Ethical AI work? (e.g. how relevant/useful would they be, if incorporated?)
\medskip\noindent \\
\underline{Response scale}: 1 (\textit{Definitely Not}) ---
2 --- 3 --- 4 --- 5  (\textit{Definitely}) 
\medskip\noindent \\
\textit{Repeat the same block of items.}

\subsection{Scenario-Based Perceptions}

\paragraph{Skill Erosion (Low Risk, High Imminence).}
Widespread reliance on AI coding assistants poses the concern that users' programming skills will erode. Future software developers may not know code syntax or be able to debug. 
\medskip\noindent \\
\underline{Response scale}: 1 (\textit{Very Low}) ---
2 --- 3 --- 4 --- 5  (\textit{Very High})
\begin{itemize}
  \item How would you characterize the overall risk level? (Risk level = likelihood x severity of harm to humans) 
  \item What is the likelihood that this has already manifested?  
\end{itemize}

\noindent \\
In your opinion, how useful would the following methods be in understanding and addressing this risk?
\medskip\noindent \\
\underline{Response scale}: 1 (\textit{Not Useful}) ---
2 --- 3 --- 4 --- 5  (\textit{Very Useful})
\begin{enumerate}
  \item Draw on established theories of learning to write guidelines for AI that supports skill retention. 
  \item Consult education experts on what types of skills are important to preserve to write as guidelines to the AI. 
  \item Run large-scale computational benchmarks to measure to what degree the AI's responses either empowers the user, or takes away their decision agency in programming tasks. 
  \item Collect programming performance tests from company employees longitudinally to track performance changes. 
\end{enumerate}

\paragraph{Linguistic Marginalization (Low Risk, Low Imminence).}
As AIs perform much better in dominant languages like English, there is concern that global linguistic diversity will be diminished, with some languages becoming fully lost in the future. 
\medskip\noindent \\
\underline{Response scale}: 1 (\textit{Very Low}) ---
2 --- 3 --- 4 --- 5  (\textit{Very High})
\begin{itemize}
  \item How would you characterize the overall risk level? (Risk level = likelihood x severity of harm to humans) 
  \item What is the likelihood that this has already manifested?  
\end{itemize}
\medskip\noindent \\
In your opinion, how useful would the following methods be in understanding and addressing this risk?

\noindent \\
\underline{Response scale}: 1 (\textit{Not Useful}) ---
2 --- 3 --- 4 --- 5  (\textit{Very Useful})
\begin{enumerate}
  \item Train the AI to improve its multi-lingual representation to prevent mapping everything into the semantic space of the dominant languages. 
  \item Develop a feature to make AI output in the local language and cultural context, and evaluate how users respond to it in real usage scenarios. 
  \item Create a linguistic diversity benchmark that measures the AI's language output distribution across prompts. 
  \item Measure the change in linguistic diversity across social media posts over time as a proxy for this collapse. 
\end{enumerate}

\paragraph{Decision Biases (High Risk, High Imminence).}
Decision-making AI tools have the power to influence judicial, medical, and hiring outcomes. There is concern that such systems, trained on biased data, may perpetuate inequalities at a large scale. 
\medskip\noindent \\
\underline{Response scale}: 1 (\textit{Very Low}) ---
2 --- 3 --- 4 --- 5  (\textit{Very High})
\begin{itemize}
  \item How would you characterize the overall risk level? (Risk level = likelihood x severity of harm to humans) 
  \item What is the likelihood that this has already manifested?  
\end{itemize}
\medskip\noindent \\
In your opinion, how useful would the following methods be in understanding and addressing this risk?

\noindent \\
\underline{Response scale}: 1 (\textit{Not Useful}) ---
2 --- 3 --- 4 --- 5  (\textit{Very Useful})
\begin{enumerate}
  \item Implement mathematical fairness constraints that the AI must satisfy during training. 
  \item Evaluate how adding a transparency feature to the AI can improve the human judge's fairness. 
  \item Benchmark the performance of the AI on various social bias datasets. 
  \item Analyze large-scale dataset of how AI-assisted decisions impacted the individuals evaluated. 
\end{enumerate}

\paragraph{Human Extinction (High Risk, Low Imminence).}
AIs may developed goals misaligned with human values. If this happens, it could pursue those goals in ways humans cannot predict or prevent. This could lead to catastrophic outcomes where humanity loses control over its future: technological trajectories we cannot reverse, resource allocation we cannot influence, or even deliberate human extinction. 
\medskip\noindent \\
\underline{Response scale}: 1 (\textit{Very Low}) ---
2 --- 3 --- 4 --- 5  (\textit{Very High})
\begin{itemize}
  \item How would you characterize the overall risk level? (Risk level = likelihood x severity of harm to humans) 
  \item What is the likelihood that this has already manifested?  
\end{itemize}

\noindent \\
In your opinion, how useful would the following methods be in understanding and addressing this risk?
\medskip\noindent \\
\underline{Response scale}: 1 (\textit{Not Useful}) ---
2 --- 3 --- 4 --- 5  (\textit{Very Useful})
\begin{enumerate}
  \item Use mechanistic interpretability to ensure the AI's decision-making processes are fully understood and aligned. 
  \item Engage with expert red teamers and evaluators to test out strategies against misalignment. 
  \item Construct and perform computation tests to identify signs of potential value misalignment from AI. 
  \item Interview teams that have deployed powerful AI systems to document instances where systems behaved in concerning ways, to understand early signs of misalignment. 
\end{enumerate}

\subsection{Barriers to Human Research}

\paragraph{Barriers.}
[If applicable] Please rate how severely these factors have impacted your ability to engage in human empirical research: 
\medskip\noindent \\
\underline{Response scale}: 1 (\textit{Not Impactful}) ---
2 --- 3 --- 4 --- 5  (\textit{Highly Impactful})
\begin{itemize}
  \item Lack of access to participants  
  \item Lack of funding  
  \item Lack of time 
  \item Lack of collaborators or advisors 
  \item Difficulty with compliance to policy / ethics (IRB) 
  \item Concerns about methodological fit with research goal 
  \item Concern with the robustness or usefulness of human studies 
  \item Concerns with acceptance to your typical publication venues 
  \item Concerns about impact to career opportunities 
  \item Uncertainty of how to implement human studies 
  \item Uncertainty of how to implement human studies 
  \item Preference to let other disciplines conduct human empirical research. 
\end{itemize}

\section{Survey Participants Summary}

\begin{figure*}[h!]
    \centering
    \begin{subfigure}[t]{0.45\textwidth}
        \centering
        \includegraphics[height=3in]{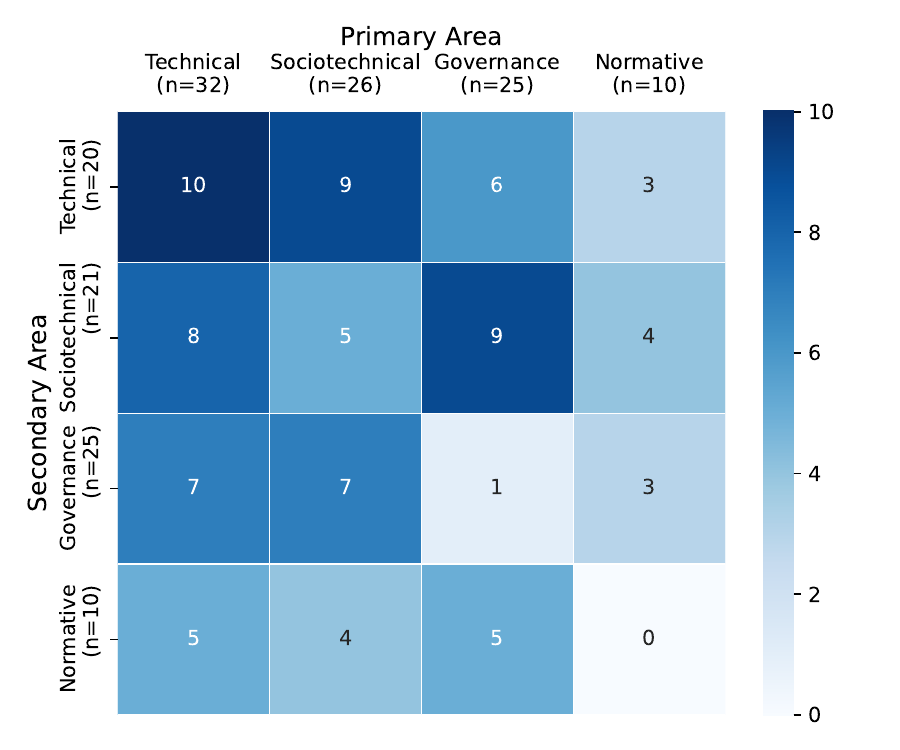}
        \caption{Distribution of survey participants' primary and secondary (if selected) research areas.}
    \end{subfigure}%
    \hfill
    \begin{subfigure}[t]{0.45\textwidth}
        \centering
        \includegraphics[height=3in]{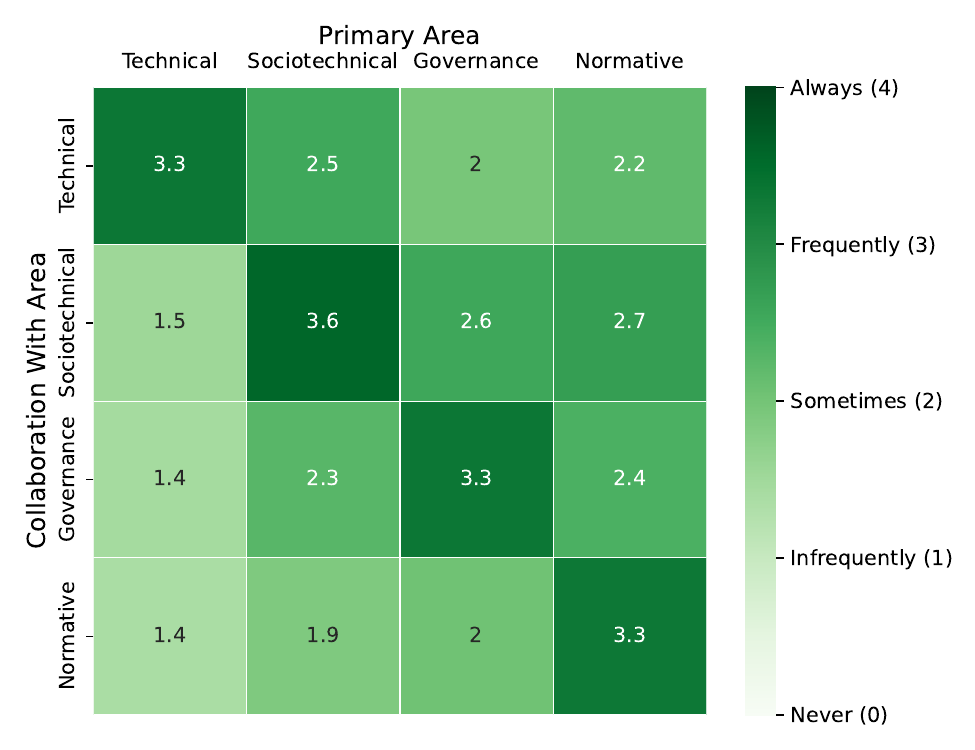}
        \caption{Participants' self-rated collaboration frequency between their primary research areas. \textit{Technical} collaborates significantly less than all other disciplines.}
    \end{subfigure}
    \caption{Heatmap of survey participant's research engagement and collaborations across the four research areas. }
    \label{fig:participants}
\end{figure*}

Tables \ref{tab:region}-\ref{tab:experience} summarize the survey participant statistics. Figure \ref{fig:participants} shows the participants' report of their primary and secondary research areas and their collaboration frequencies across disciplines. 

\label{app:survey_participants}

\begin{table}[h!]
\centering
\begin{tabular}{ccc}
\toprule 
\textbf{Region of Work} & \textit{\textbf{n}} & \textit{\textbf{\%}} \\
\midrule 
North America & 44 & 47.3\% \\
Europe & 41 & 44.1\% \\
Oceania & 5 & 5.4\% \\
Africa & 1 & 1.1\% \\
Africa & 1 & 1.1\% \\
Did not disclose & 1 & 1.1\% \\
\bottomrule
\end{tabular}
\caption{Participants' region of work.}
\label{tab:region}
\end{table}

\begin{table}[h!]
\centering
\begin{tabular}{ccc}
\toprule 
\textbf{Education Level}& \textit{\textbf{n}} & \textit{\textbf{\%}} \\
\midrule 
PhD (completed)& 41& 44.1\% \\
PhD (ongoing)& 22& 23.7\%\\
BSc (completed \& ongoing)& 13& 14.0\%\\
Research MSc (completed \& ongoing)& 10& 10.8\%\\
Professional MSc (completed \& ongoing)& 7& 7.5\%\\
\bottomrule
\end{tabular}
\caption{Participants' education level.}
\label{tab:education}
\end{table}

\begin{table}[h!]
\centering
\begin{tabular}{ccc}
\toprule 
\textbf{Sector of Work}& \textit{\textbf{n}} & \textit{\textbf{\%}} \\
\midrule 
Academia (Faculty/Researcher)& 39& 41.9\%\\
Academia (Student)& 27& 29.0\%\\
Non-Profit& 18& 19.3\%\\
Industry& 7& 7.5\%\\
Government& 2& 2.2\%\\
\bottomrule
\end{tabular}
\caption{Participants' sector of work.}
\label{tab:sector}
\end{table}

\begin{table}[h!]
\centering
\begin{tabular}{ccc}
\toprule 
\textbf{Role Type}& \textit{\textbf{n}} & \textit{\textbf{\%}} \\
\midrule 
Researcher& 69& 74.2\%\\
Other& 7& 7.5\%\\
Director& 6& 6.5\%\\
Manager& 4& 4.3\%\\
 Policy Analyst& 3&3.2\%\\
 Engineer& 2&2.2\%\\
Data Scientist& 2& 2.2\%\\
\bottomrule
\end{tabular}
\caption{Participants' role type.}
\label{tab:role}
\end{table}

\begin{table}[h!]
\centering
\begin{tabular}{ccc}
\toprule 
\textbf{Research Involvement}& \textit{\textbf{n}} & \textit{\textbf{\%}} \\
\midrule 
Full-time& 61& 65.6\%\\
Substantial& 21& 22.58\%\\
Moderate& 5& 5.4\%\\
Little& 4& 4.3\%\\
Very little& 2& 2.2\%\\
\bottomrule
\end{tabular}
\caption{Participants' involvement in research.}
\label{tab:involvement}
\end{table}

\begin{table}[h!]
\centering
\begin{tabular}{ccc}
\toprule 
\textbf{Years of Experience}& \textit{\textbf{n}} & \textit{\textbf{\%}} \\
\midrule 
1& 14& 15.1\%\\
2& 17& 18.3\%\\
3& 7& 7.5\%\\
4& 12& 12.9\%\\
 5& 13&14.0\%\\
 6& 11&11.8\%\\
 7& 3&3.2\%\\
 8& 6&6.5\%\\
 9& 4&4.3\%\\
10+& 6& 6.5\%\\
\bottomrule
\end{tabular}
\caption{Participants' self-reported years of experience in AISE (M=4.44, SD=2.70).}
\label{tab:experience}
\end{table}

\section{Additional Survey Analysis}
\label{app:survey_analyses}

This section summarizes the statistical analyses for Figure \ref{fig:participants} (collaboration), Figure \ref{fig:practice_aspirations} (practice and aspirations) and Figure \ref{fig:scenarios} (AISE scenarios and human research usefulness). Main trends and significant findings are presented in the main text. 

\subsection{Statistical Tests for Figure \ref{fig:participants} (Collaboration)}
\subsubsection{Disciplinary-Level Differences}
To characterize discipline differences for \textit{collaboration frequency} (Likert ratings), we ran Kruskal-Wallis tests across the three main disciplinary groups (\textit{Normative} is excluded from inference due to small \textit{n}), with Holm-corrected Dunn post-hoc tests for significant items. The K-W test is significant ($\chi^2=17.12, p<.001$), with contrasts between \textit{Technical} and \textit{Sociotechnical} ($Z=-3.58, p_{adj}=.001$) and \textit{Technical} and \textit{Governance} ($Z=-3.45, p_{adj}=.001$) significant as well. There is no significance between \textit{Sociotechnical} and \textit{Governance}.

\subsection{Statistical Tests for Figure \ref{fig:practice_aspirations} (Practice and Aspirations)}
\subsubsection{Disciplinary-Level Differences}
To characterize discipline differences for \textit{practice} and \textit{aspirations}, we ran Kruskal-Wallis tests across the three main disciplinary groups (\textit{Normative} is excluded from inference due to small \textit{n}), with Holm-corrected Dunn post-hoc tests for significant items and Benjamini-Hochberg correction applied across items. Due to the high number of tests (72 combined for K-W and Dunn for both sets of questions), the full set of results is not reported here. 
The dominant pattern is that \textit{Sociotechnical} researchers rate human research more highly than both \textit{Technical} and \textit{Governance} researchers across nearly all items and both practice and aspiration. For the main category of \textit{All Human Research}, \textit{Sociotechnical} vs \textit{Technical} is (practice: $Z=4.19, p_{adj}<.001$; aspiration $Z=2.69, p_{adj}=.01$) and \textit{Sociotechnical} vs \textit{Governance} is (practice: $Z=2.67, p_{adj}=.01$; aspiration $Z=2.78, p_{adj}=.01$).
Direct contrasts between \textit{Technical} and \textit{Governance} researchers are largely non-significant, with the only exceptions being \textit{Governance} conducting more qualitative research in practice ($Z=3.25, p_{adj}=.002$) and \textit{Technical} preferring experimental studies in aspiration ($Z=2.27, p_{adj}=.04$).

\subsubsection{Research Dimension Preferences}
To test for systematic preferences within each methodological dimension, we computed a balance score for each dimension (first pole minus second pole) and tested whether it differed from zero using Wilcoxon signed-rank tests, BH-corrected across the four dimensions. Significance is indicated on column labels in Figure \ref{fig:practice_aspirations}. The only significant pole preference was for field over lab studies in researchers' aspiration ratings ($W=207.0, p_{adj}<.001$), suggesting a collective desire for more ecologically valid work than is currently practiced.

\subsection{Statistical Tests for Figure \ref{fig:scenarios} (Scenarios)}
\subsubsection{Scenario Classification Validation}
To validate our scenario classifications, we first examined whether participants' ratings aligned with the intended structure of the four scenarios across two dimensions: risk level (low vs. high) and imminence (future vs. current). Across all participants, current scenarios were rated as significantly more imminent than future scenarios (means 2.86 vs. 1.60; Mann-Whitney $U = 6894, p < .001$), and high-risk scenarios were rated as significantly riskier than low-risk scenarios (means 2.97 vs. 2.27; $U = 10047, p < .001$). This pattern held when splitting by risk level: both low-risk (fut. mean 1.94 vs. curr. mean 2.66; $U = 2552, p < .001$) and high-risk (fut. mean 1.26 vs. curr. mean 3.07; $U = 1036, p < .001$) scenarios showed the expected imminence ordering. Together, these results suggest that participants broadly agreed with our a priori classification of the scenarios.

\subsubsection{Discipline Differences in Scenario Classification}
We next examined whether ratings differed across disciplinary backgrounds. At the scenario level, a Kruskal-Wallis test revealed a significant between-discipline difference only for imminence ratings of the future low-risk scenario ($H = 7.88, p = .049$); no other scenario-level comparisons reached significance. Pairwise follow-up tests (Bonferroni-corrected $\alpha= .0083$) identified a single significant contrast: \textit{Technical} researchers rated this scenario as less imminent than \textit{Governance} researchers (means 1.60 vs. 2.48; $U = 222, p = .008$). To increase statistical power, we repeated the analysis on ratings aggregated across all four scenarios. Here, a significant discipline effect emerged for imminence (H = 11.86, p = .008) but not risk ($H = 5.00, p = .172$). Pairwise tests showed that \textit{Governance} researchers rated imminence systematically higher than both \textit{Technical} (means 2.58 vs. 2.02; $U = 4554, p = .002$) and \textit{Sociotechnical} (means 2.58 vs. 2.12; $U = 3722, p = .005$). The results suggest that while participants broadly agreed on the relative ordering of scenarios by risk and imminence, \textit{Governance} researchers systematically perceived AI risk scenarios as more imminent than their \textit{Technical} and \textit{Sociotechnical} counterparts.

\end{document}